\documentclass{dhbenelux}

\usepackage{booktabs} 
\usepackage{authblk} 
\usepackage{graphicx} 
\usepackage{amsmath} 
\usepackage{subcaption}
\usepackage{multirow}

\author{Mhd Hussein Murtada}
\author{Z. Faidon Brotzakis}
\author{Michele Vendruscolo \footnote{Correspondence: mv245@cam.ac.uk} }

\affil{\textit{Centre for Misfolding Diseases, Yusuf Hamied Department of Chemistry, 
University of Cambridge, Cambridge, UK}}

\title{MD-LLM-1: A Large Language Model for Molecular Dynamics}

\begin{document}

\maketitle

\begin{abstract}
    \noindent {Molecular dynamics (MD) is a powerful approach for modelling molecular systems, but it remains computationally intensive on spatial and time scales of many macromolecular systems of biological interest. To explore the opportunities offered by deep learning to address this problem, we introduce a Molecular Dynamics Large Language Model (MD-LLM) framework to illustrate how LLMs can be leveraged to learn protein dynamics and  discover states not seen in training. By applying MD-LLM-1, the first implementation of this approach, obtained by fine-tuning Mistral 7B, to the T4 lysozyme and Mad2 protein systems, we show that training on one conformational state enables the prediction of other conformational states. These results indicate that MD-LLM-1 can learn the principles for the exploration of the conformational landscapes of proteins, although it is not yet modeling explicitly their thermodynamics and kinetics.

}
\end{abstract}

\newpage
\section{Introduction}

Proteins are dynamic molecules whose function is intimately linked with their ability to sample different conformational states\cite{Benkovic2003, Mittermaier2006, Vendruscolo2006, HenzlerWildman2007, Boehr2009}. Since protein motions underlie most biological processes, the ability to characterise them is crucial in a wide range of applications\cite{Benkovic2003, Mittermaier2006, Vendruscolo2006, HenzlerWildman2007, Boehr2009}. 
\\
\\
As many functional motions of proteins involve the exploration of the conformational space under effectively equilibrium conditions, several approaches have focused on the generation of conformational ensembles corresponding to the Boltzmann distribution\cite{Bonomi2016, Bussi2020}. Based on the remarkable success of machine learning (ML) methods in protein structure predictions\cite{Jumper2021, abramson2024n, Krishna2024}, one can ask whether the exploration of the structural ensembles of proteins could be implemented using some forms of ML\cite{noé2019, von2025machine, lewis2025scalable, jing2024alphafold, brotzakis2025alphafold, schnapka2025atomic, aranganathan2025modeling}.
\\
\\
In many applications, one is specifically interested in following the dynamics of proteins according to the physical laws of motion. Over the last 60 years, molecular dynamics (MD) simulations have represented the gold standard for this purpose, as they provide atomic-level detail by implementing the exploration of the conformational space of proteins by integrating numerically the equations of motion\cite{alder1957phase, McCammon1977, Karplus2002, car1985unified}. However, achieving convergence in MD simulations often requires extensive computational resources and time\cite{piana2011bj, vendruscolo2011protein}, or the application of system-specific enhanced sampling methods\cite{Lindorff-Larsen2005, Bussi2020, Bonomi2016}, particularly when exploring rare events or high-energy states\cite{sarich2013e, ray2020weighted, bolhuis2002transition, brotzakis2021method}. This computational burden limits the current ability to study long-timescale processes and rare conformational transitions. 
\\
\\
Addressing this problem using ML offers novel opportunities to combine the efficiency of deep learning methods with the physical accuracy of MD simulations. A successful approach, however, ought to balance multiple competing demands: adherence to physical laws, achievement of computational efficiency and accuracy in capturing conformational diversity. Methods to address this problem have been proposed based on the use of ML to speed up the calculations of force fields \cite{Behler2007, Behler2016, Noe2020, batatia2025design} or to implement long time steps\cite{schreiner2023implicit, bigi2025flashmd}.
\\
\\
Since language models (LMs) are powerful tools for pattern recognition and generation\cite{Vaswani2017, zhao2023survey}, they may offer further avenues to address the challenges involved in generating MD trajectories that capture the underlying laws of motion. LMs have already been exploited in a range of applications in protein science\cite{xiao2025protein}, including for protein structure prediction\cite{lin2023evolutionary} and protein design\cite{madani2023large}. While earlier studies demonstrated the potential of using LMs in molecular dynamics through recurrent neural networks (RNN) and long short-term memory (LSTM) architectures for small systems\cite{tsai2020learning},  recent advances leading to the development of large language models (LLMs) present expanded possibilities for modeling complex biomolecular systems.
\\
\\
In this work, we build on our initial report on the use of LMs for MD\cite{murtada2024language} to explore the application of state-of-the-art LLMs to molecular dynamics through a Molecular Dynamics Large Language Model (MD-LLM) framework. By leveraging the Mistral 7B architecture\cite{jiang2024mistral} fine-tuned with Low-Rank Adaptation (LoRA)\cite{hu2021lora}, we report the first implementation of this framework (MD-LLM-1). Our results indicate that MD-LLM-1 can discover low population states not seen during training. The transformer architecture underlying these models offers parallel processing capabilities and effective modeling of long-range dependencies\cite{Vaswani2017}, enabling the discovery of cross-state transitions in biologically relevant proteins. Their self-attention mechanism\cite{Vaswani2017} allows the model to simultaneously evaluate relationships between different parts of the protein structure and learn the underlying principles governing conformational changes over time.
\\
\\
We illustrate the use of MD-LLM-1 through its application to two well-characterized proteins, T4 lysozyme\cite{bouvignies2011solution} and Mad2\cite{JainSekhar2025}. We chose T4 lysozyme since the native and excited state structures of two of its mutational variants have been experimentally determined using nuclear magnetic resonance (NMR) spectroscopy\cite{bouvignies2011solution}. The L99A mutant exhibits a native state population of 97\% and an excited state population of 3\%, while a triple mutation (L99A, G113A, R119P) inverts these populations, with the excited state of L99A becoming the native state of the triple mutant (96\%) \cite{bouvignies2011solution}. Through LLM fine-tuning and prompt engineering, we show that our approach achieves cross-state discovery capabilities. An MD-LLM-1 trained exclusively on the native state of the L99A mutant samples conformations characteristic of the excited state of this mutant, while an MD-LLM-1 trained on the native state of the triple mutant can predict the native state conformation of the L99A mutant. For Mad2, we show that training on one major conformational state enables the discovery of the other major state. These findings illustrate the potential of the MD-LLM framework for exploring conformational landscapes of proteins.

\section{Methods}
\subsection{Overview of the methodological framework}
MD-LLM-1 consists of three integrated components:
\begin{enumerate}
    \item\textbf{System-specific learning}: We fine-tune Mistral 7B\cite{jiang2024mistral} using Low-Rank Adaptation (LoRA)\cite{hu2021lora} on short MD trajectories containing conformations from a single state (e.g., native state conformations for T4 lysozyme). As training data, we use a trajectory of encoded conformations structured in a rolling window fashion where each set of N consecutive frames predicts the following frame. A specialized prompt template guides the model to understand the relationship between sequential protein conformations and capture the underlying physical patterns.
    \item\textbf{Conformational state discovery}: We use the fine-tuned MD-LLM-1 to discover conformational states through sequential inference. By running inference starting from the training trajectory, we enable the exploration of the conformational space not present in the original training data. This process leverages the understanding of the model of protein dynamics to sample low population states.
    \item\textbf{Structure decoding and ensemble generation}: The tokens generated by MD-LLM-1 are decoded back to three-dimensional protein coordinates using the FoldToken decoder network\cite{gao2024foldtoken4}. The stochastic nature of the decoder enables the efficient generation of conformational ensembles, providing a computationally efficient alternative to running conventional MD simulations.
\end{enumerate}

\subsection{Structural representation}
In the MD-LLM-1 approach, protein structures are represented using the FoldToken tokenization scheme \cite{gao2024foldtoken4}. This method transforms complex 3D conformations of proteins into sequences of discrete numerical tokens that can be processed by LLMs.
The FoldToken process involves three main steps \cite{gao2024foldtoken4}:
\begin{enumerate}
    \item \textbf{Protein graph representation:} The protein structure is represented as a graph where each node represents a residue and each edge represents a spatial relationship.
    \item \textbf{Feature extraction:} A BlockGAT\cite{gao2024foldtoken4, gao2024uniif} encoder processes this graph to extract invariant structural features:
    $$
    f_i = \text{BlockGAT}(G(\{B_s\}_{s=1}^n, E))
    $$
    where $f_i$ is the embedding of the i-th residue.
    \item \textbf{Vector quantization:} The continuous embeddings are converted to discrete tokens through a quantization function:
    $$
    z_i = Q(f_i)
    $$
    The function Q maps each continuous embedding to its nearest representative in a learned codebook, assigning a discrete token ID \cite{gao2024foldtoken4}. This process effectively compresses the complex structural information into a form that can be processed by LLMs.
\end{enumerate}
For example, since T4 lysozyme is a 164-residue protein, each conformation is represented as a sequence of 164 numerical tokens. These tokens encode the structural information including backbone atom positions, relative orientations, and torsional relationships in a format compatible with LLM processing.
\\
\\
When generating new structures, the numerical tokens are mapped back to structural embeddings through a decoder:
$$
X_{\text{pred}} = \text{Decoder}([z_1, z_2, \ldots, z_n])
$$
where $X_{pred}$ represents the predicted 3D coordinates of the protein structure. The decoder consists of an SE(3)-equivariant neural network that transforms the token embeddings back into three-dimensional coordinates, ensuring that the generated structures adhere to physical constraints \cite{gao2024foldtoken4,gao2024uniif}.
\\
\\
This protein representation is well suited for LLMs for several reasons. First, its discrete token space maps to the vocabulary-based processing paradigm of LLMs, allowing Mistral 7B\cite{jiang2024mistral} to process protein conformations as if they were text. Second, the numerical tokens create a consistent encoding scheme that the LLM can learn to predict sequentially, enabling the semantic modeling of conformational transitions over time. Third, the compressed representation significantly reduces the dimensionality of the protein structural space \cite{gao2024uniif}, making it feasible to learn the mapping between sequential conformations with limited training data.
\\
\\
In our framework, we effectively repurpose FoldToken for temporal and semantic modeling of protein dynamics by arranging tokens in a time-sequential manner. Specifically, we format the data as sequences of N consecutive frames to predict the (N+1) frame, enabling the model to learn the temporal evolution of protein conformations. This numerical token representation, structured to handle sequences of conformations across time, serves as the foundation for our prompt-based approach in MD-LLM-1, where the model learns to generate the natural progression of protein motion.

\subsection{Fine-tuning Mistral 7B with LoRA}
MD-LLM-1 is fine-tuned using Low-Rank Adaptation (LoRA) \cite{hu2021lora} for the specific task of MD predictions. We selected this approach due to its ability to adapt large pre-trained models to specialized domains with minimal computational overhead while maintaining performance.

\subsubsection{Model architecture and optimisation}
We fine-tuned the mistral-7b-v0.3-bnb-4bit \cite{unsloth_mistral7b_2024} quantised variant from the Unsloth\cite{unsloth} HuggingFace repository.
\\
\\
\textbf{Model architecture.} The model architecture has the following components:
\begin{itemize}
    \item \textbf{Base architecture}: MistralForCausalLM with $L=32$ transformer layers
    \item \textbf{Dimensionality}: Hidden dimension $d_{\text{model}}=4096$ with feed-forward dimension $d_{\text{ff}}=14336$
    \item \textbf{Attention}: Multi-head attention with $h=32$ heads, using grouped-query attention with $h_{\text{kv}}=8$ key-value heads
    \item \textbf{RoPE}: Rotary positional embeddings\cite{su2024roformer} applied with base frequency $\theta=1,000,000$, where position $m$ for dimension $i$ is encoded as:
    $$\begin{pmatrix} \cos(m\theta^{-2i/d}) & -\sin(m\theta^{-2i/d}) \\ \sin(m\theta^{-2i/d}) & \cos(m\theta^{-2i/d}) \end{pmatrix}$$
    \item \textbf{Context length}: Maximum sequence length $L_{\text{max}}=32,768$ tokens
    \item \textbf{Precision}: 4-bit quantization with bfloat16 compute precision
\end{itemize}
The RoPE mechanism\cite{su2024roformer} is particularly valuable for MD as it encodes relative positional information directly in the attention mechanism, enabling better modeling of spatial relationships between residues across sequential frames.
\\
\\
\textbf{Fine-tuning optimisation.}
We employed LoRA (Low-Rank Adaptation)\cite{hu2021lora} for efficient fine-tuning. LoRA works by introducing trainable low-rank matrices to the pre-trained weights according to:
$$
W = W_0 + \Delta W = W_0 + BA
$$
where $W_0 \in \mathbb{R}^{d \times k}$ represents the frozen pre-trained weights, and $B\in \mathbb{R}^{d \times r}$ and $A\in \mathbb{R}^{r \times k}$ are low-rank matrices with $r \ll \min(d, k)$ is the rank parameter that determines the dimensionality of the low-rank matrices. This approach significantly reduces the number of trainable parameters compared to full fine-tuning while preserving the core capabilities of the model\cite{hu2021lora}.
\\
\\
The performance of our training process, powered by the Unsloth framework \cite{unsloth}, was substantially enhanced by Flash Attention 2 \cite{dao2023flashattention}, a memory-efficient attention implementation that reduces the memory complexity of the self-attention mechanism from O(N²) to O(N), where N is the sequence length \cite{dao2023flashattention}. This optimization was particularly valuable for our application, as it allowed us to process long sequences of protein conformational states with reduced memory overhead. Flash Attention achieves this efficiency by breaking the attention computation into blocks that fit in fast GPU memory, eliminating the need to store the full attention matrix and intermediate results in high-bandwidth memory \cite{dao2023flashattention}.
\\
\\
Additional optimizations from the Unsloth framework \cite{unsloth} included Triton-based kernels \cite{hsu2024liger} that replaced generic PyTorch operations, efficient weight quantization that maintained 4-bit precision for most parameters while selectively upcasting only when necessary for accurate computations, and memory-optimized implementation of cross-entropy loss. Collectively, these optimizations enabled us to fine-tune the 7 billion parameter model efficiently on standard research hardware.

\subsubsection{LoRA configuration and key model components}
We configured LoRA with a rank of 16 and applied it to the critical projection matrices within the transformer architecture of the model. These matrices play distinct roles in processing the complex patterns of protein motion. The following is a list of the target matrices that were fine tuned with LoRA:
\\
\\
In the attention mechanism:
\begin{itemize}
    \item \textbf{Query projections}: Transform input embeddings into query representations that seek relevant information, enabling the model to identify correlations between specific residue positions across consecutive frames.
    \item \textbf{Key projections}: Create key representations that encode the information content of each position, allowing the model to index important conformational features.
    \item \textbf{Value projections}: Generate value representations containing the actual information to be aggregated, holding the structural details that will be selectively combined.
    \item \textbf{Output projections}: Transform the aggregated attention outputs back to the dimension space of the model, integrating the attended information into the ongoing representation.
\end{itemize}
In the feed-forward network:
\begin{itemize}
    \item \textbf{Gate projections}: Control information flow in the SwiGLU activation function, acting as adaptive filters that determine which conformational features should be emphasized or suppressed.
    \item \textbf{Up projections}: Expand the representation dimension, creating a higher-dimensional space where complex non-linear patterns in protein dynamics can be modeled.
    \item \textbf{Down projections}: Compress the information back to the dimension of the model, distilling the essential features captured in the expanded representation.
\end{itemize}
Targeting these specific projection matrices with LoRA is important for adapting the model to protein conformational dynamics. The attention mechanism components enable the model to identify correlations between residue positions across time frames, which is essential for understanding how local structural changes propagate through the protein \cite{Vaswani2017}. Meanwhile, the feed-forward components allow the model to capture complex non-linear relationships in the conformational space, which are critical for modeling transitions between different states of the protein.
\\
\\
The fine-tuning process with a scaling factor $\alpha$ of 16 and no dropout maximized computational efficiency while providing sufficient capacity to learn the specialized MD patterns.

\subsubsection{Training parameters and dataset preparation}
We structured the training data from a short MD trajectory containing only conformations of one state among those populated by the protein under study. Each training example followed an overlapping window approach, where frames 1-10 were used to predict frame 11, frames 11-20 to predict frame 21, and so on. This structure enabled the model to learn the temporal relationships between sequential conformational states.
\\
\\
The LLM interface was defined through a set of prompts that provided clear instruction on the task requirements. This prompt engineering approach gave the model clear context about the task while enforcing constraints on the output format, ensuring that predictions maintained the exact structure required for subsequent processing steps in the MD-LLM-1 pipeline.
\\
\\
The training process was conducted on NVIDIA A100 GPU using a batch size of 2 per device with 4 gradient accumulation steps (effective batch size of 8), a learning rate of 2e-4 with linear scheduling, and the AdamW 8-bit optimizer for memory efficiency. We maintained a maximum sequence length of 9000 tokens to accommodate the full context needed for protein trajectory analysis, with mixed precision (Bfloat16).

\subsection{Inference for sequential conformation generation}
After fine-tuning the Mistral 7B model, we used it to extend trajectories through sequential inference beyond the original training data length. This trajectory extension process enables the discovery of conformational states not present in the training data, effectively allowing the model to explore regions of conformational space that were kinetically inaccessible during the original simulation. 

\subsubsection{Sequential frame generation}
Our inference approach utilized a rolling window strategy where N consecutive frames of conformations were used to predict the next frame in the sequence. The process began with the last N frames from the training data to ensure continuity in the protein's motion and then proceeded iteratively through extended inference, enabling exploration beyond the original conformational space:
$$
F_{n+1}=MD{\text -}LLM{\text -}1(F_{n-9},F_{n-8},…,F_{n})
$$
where $F_i$ represents the conformation frame at step $i$, and $MD{\text -}LLM{\text -}1$ is our fine-tuned model that predicts the next frame based on the previous 10 frames.
\\
\\
For each prediction step, we employed controlled sampling parameters to maintain diversity while ensuring structural plausibility:
\begin{itemize}
    \item Temperature of 1.0 to introduce appropriate stochasticity
    \item $top\_k=100$ and $top\_p=0.95$  sampling creates a balanced approach to token selection. Top-k sampling first limits the possible options to the 100 most probable tokens, while nucleus sampling (top-p) further refines selection to the smallest subset of tokens whose cumulative probability reaches 95\%. This dual-constraint mechanism ensures generated frames maintain a balance between conformational diversity and structural coherence by preventing both highly improbable token selections and overly deterministic predictions.
    \item Single beam search to efficiently generate diverse conformations \cite{Moret2021}
\end{itemize}
These parameters enabled our model to generate trajectories that could explore conformational states not present in the training data, facilitating the discovery of low population states through the learned understanding of the model of protein dynamics. For each position $i$, the model output is a probability distribution over possible tokens in vocabulary:
$$
P(token_i | token_{<i})) = \text{softmax}\left(\frac{logits_i}{T_{\text{sampling}}}\right)
$$
where $T_{\text{sampling}}=1.0$ scales the logits to control sampling diversity. The beam search \cite{Moret2021} maintains the top-k partial sequences by score:
$$
\text{score}(b) = \sum_{i=1}^{l} \log P(token_i^b | token_{<i}^b)
$$

\section{Results}
\subsection{Generation of protein conformations}
Despite having no explicit knowledge of the chemistry and physics of proteins, or of molecular mechanics, and being pretrained solely on human language, the fine-tuned Mistral 7B model generates physically valid protein conformations that are structurally similar to the training data. To assess the fundamental capability of the model to learn protein structural principles, we evaluated its ability to generate novel conformations that maintain high similarity (in terms of low root mean square deviation, RMSD) when using the starting structure as a reference.
\\\\
Conformations generated by MD-LLM-1 are in agreement with the training data for both T4 lysozyme and Mad2, producing novel structures that maintained structural fidelity comparable to the original training conformations. RMSD time series revealed that generated conformations maintained high structural quality, illustrating a capability of learning of the characteristic structural features without any explicit enforcement of physical constraints such as bond lengths, angles, or molecular mechanics force fields.
\\\\
Analysis of the conformational space sampled by the MD-LLM-1 reveals both the reproduction of training-like states and the discovery of novel conformational regions. MD-LLM-1 samples conformations similar to the training data while also exploring states not present in the original training trajectory. This capability suggests that MD-LLM-1 learned general principles that enable the exploration of diverse conformational regions, including states that were not visited during the original simulation timescales, as described in the following sections.

\subsection{Cross-state discovery in T4 lysozyme}
To show the ability of MD-LLM-1 to discover conformational states beyond its training data, we evaluated it on T4 lysozyme, for which the L99A mutant and the L99A-G113A-R119P triple mutant exhibit different native states \cite{bouvignies2011solution}. The single mutant adopts native state (97\% population) and an excited state (3\% population), while the native state of the triple mutant (96\% population) corresponds to the excited state of the L99A mutant, effectively inverting the conformational equilibrium\cite{bouvignies2011solution}.
\\
\\
We conducted two complementary experiments to assess bidirectional cross-state discovery capabilities: MD-LLM-1 trained exclusively on relaxed state conformations and another MD-LLM-1 trained exclusively on excited state conformations. Both calculations illustrate the ability of MD-LLM-1  to discover alternative conformational states not present in the training data.
\\\\
\subsubsection{Discovery of excited states from native state training}
\begin{figure}[h!]
    \centering
    \begin{subfigure}[b]{0.45\textwidth}
        \centering
        \includegraphics[width=\textwidth]{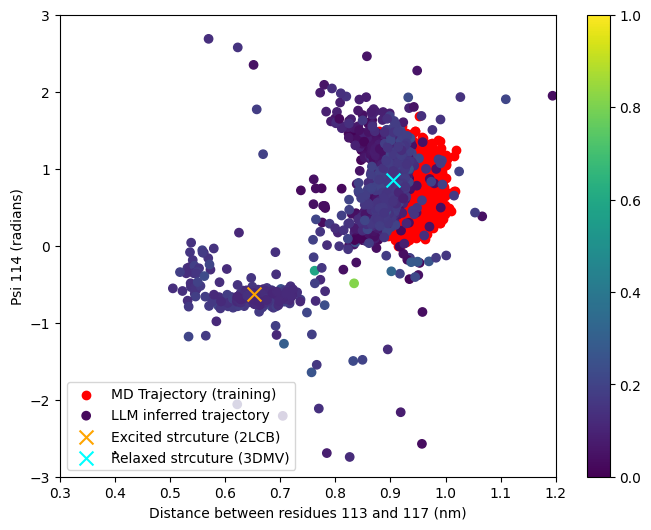}
        \caption{}
        \label{fig:t4_relaxed_scatter}
    \end{subfigure}
    \hfill
    \begin{subfigure}[b]{0.45\textwidth}
        \centering
        \includegraphics[width=\textwidth]{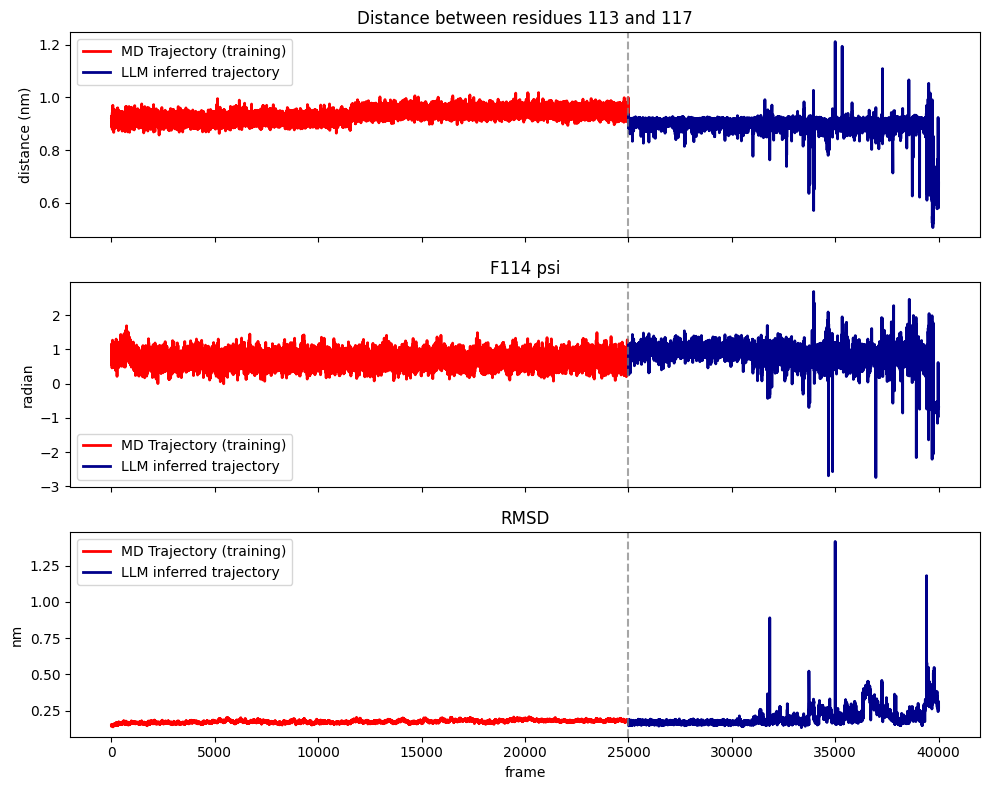}
        \caption{}
        \label{fig:t4_relaxed_ts}
    \end{subfigure}
    \caption{(a) Comparison of the regions of the conformational landscape explored by the training data (a short MD trajectory of the L99A mutant of T4 lysozyme) and by the inference of MD-LLM-1; the color bar corresponds to the RMSD from the native state. (b) Time series of characteristic distance between residues 113-117, the $\psi$ angle of residue 114, and the RMSD starting from the training data; these parameters indicate how the MD-LLM-1 inferred trajectory extrapolates the training trajectory.}
    \label{fig:t4-excited}
\end{figure}
MD-LLM-1 trained exclusively on the native state of the L99A mutant discovered excited state conformations. The training data clustered tightly around the native state of the L99A mutant (PDB: 3DMV), with the characteristic distance between residues 113 and 117 of approximately 0.9 nm and the ψ angle of residue 114 around +0.75 radians, consistent with the experimental native state values of 7 ± 1.5 Å for the 113-117 distance and 0.75 ± 0.375 radians for the ψ angle \cite{Smith2020} (Figure \ref{fig:t4_relaxed_scatter}).
\\
\\
During inference, MD-LLM-1 explored a broader conformational space, discovering a cluster of states with the 113-117 distance as short as 0.5 nm and the ψ angle reaching -0.75 radians (Figure \ref{fig:t4_relaxed_scatter}). These discovered conformations cluster around the experimental structure of the excited state (PDB: 2LCB), showing the ability of MD-LLM-1 to identify conformational states it had never encountered during training. Other conformations are scattered in the conformational landscape that may correspond to intermediate structures.
\\
\\
Time series analysis reveals the transition from training reproduction to state discovery (Figure \ref{fig:t4_relaxed_ts}). The initial part (in red) shows stable parameters consistent with the relaxed state training data, while the MD-LLM-1 inference part (in dark blue) exhibits exploration of conformational space, including transitions to excited state parameters. An RMSD analysis indicates that while the discovered states explore new conformational regions, they maintain a structural similarity to known protein conformations.

\subsubsection{Discovery of the native state from the excited state training}
\begin{figure}[h!]
    \centering
    \begin{subfigure}[b]{0.45\textwidth}
        \centering
        \includegraphics[width=\textwidth]{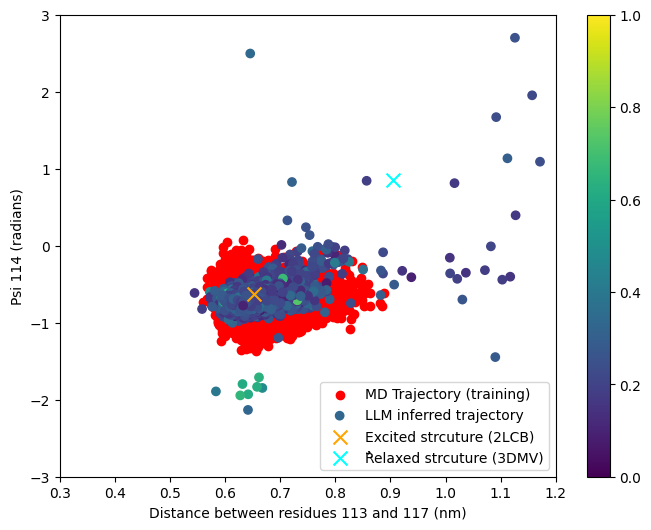}
        \caption{}
        \label{fig:t4_excited_scatter}
    \end{subfigure}
    \hfill
    \begin{subfigure}[b]{0.45\textwidth}
        \centering
        \includegraphics[width=\textwidth]{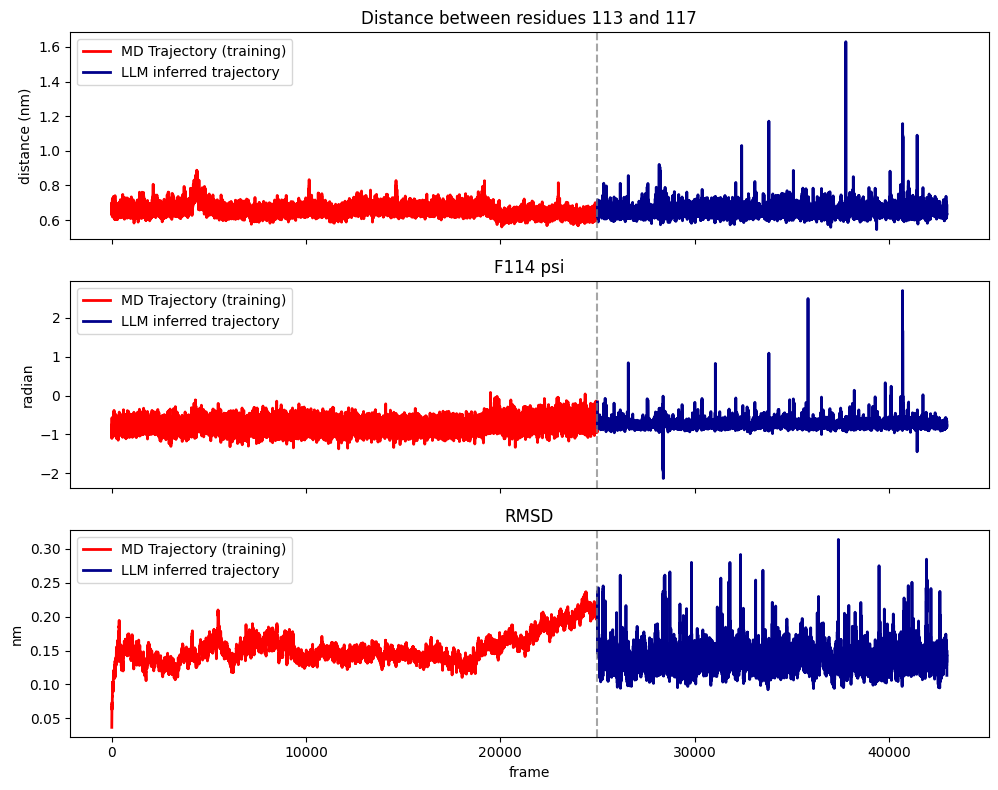}
        \caption{}
        \label{fig:t4_excited_ts}
    \end{subfigure}
    \caption{(a) Comparison of the regions of the conformational landscape of the training data (a short MD trajectory of the L99A-G113A-R119P triple mutant of T4 lysozyme) and the inferred states by MD-LLM-1; the color bar corresponds to the RMSD from the excited state. (b)  Time series of the characteristic distance between residues 113-117, the $\psi$ angle of residue 114, and the RMSD starting from the training data; these parameters indicate how the MD-LLM-1 inferred trajectory extrapolates the training trajectory.}
    \label{fig:t4-excited}
\end{figure}
The MD-LLM-1 trained exclusively on the native state of the L99A-G113A-R119P triple mutant, which correspond to the excited state of the L99A mutant, discovered native state conformations characteristic of the native state of the L99A mutant. The training data clustered around the excited state of the L99A mutant (PDB: 2lCB), with the distance between residues 113 and 117 of approximately 0.6 nm and the ψ angle of residue 114 around -0.75 radians (Figure \ref{fig:t4_excited_scatter}).
\\
\\
During extended inference, MD-LLM-1 explored conformational space toward the native state region of the L99A mutant, discovering states with the 113-117 distance extending to 0.9-1.2 nm and the 114 ψ angle reaching +3 radians (Figure \ref{fig:t4_excited_scatter}). Some of these discovered conformations fall close to the native state of the L99A mutant (PDB: 3DMV), showing the ability of MD-LLM-1 to predict the alternative conformational state not seen during training. As in the previous test, we see also other low RMSD conformations away from the expected clusters.
\\
\\
Time series analysis shows the exploration of MD-LLM-1 from reproduction of the state corresponding to the training data to the discovery of states not seen during the training (Figure \ref{fig:t4_excited_ts}). The training portion maintains excited state parameters (0.6 nm distance, -0.75 radians ψ), while the MD-LLM-1 inference portion exhibits exploration toward longer 113-117 distances and positive 114 ψ angles characteristic of the relaxed state. Notably, the RMSD values remain below 0.3 nm throughout the discovery process, indicating that the model maintains high structural quality while exploring new conformational states. This low RMSD range shows that the discovered relaxed states preserve a protein-like structure despite representing significant conformational changes from the excited state training data.

\subsubsection{Bidirectional cross-state discovery}
This bidirectional cross-state discovery capability indicates that MD-LLM-1 learned fundamental conformational relationships rather than memorizing specific structural patterns. The importance of these discoveries is highlighted by the known difficulty of sampling these low population states. Literature reports indicate high kinetic barriers between the native and excited states, with the L99A  mutant showing only 3\% excited state population and the triple mutant requiring multiple stabilizing mutations to achieve a 96\% population of the excited state of the L99A mutant \cite{bouvignies2011solution}.
\\
\\
These results establish that MD-LLM-1 can perform cross-state discovery, enabling the prediction of alternative conformational states from limited training data representing only one state of a multi-state system. 

 \subsection{Discovery of conformational states of Mad2}
 \begin{figure}[h!]
    \centering
    \begin{subfigure}[b]{0.45\textwidth}
        \centering
        \includegraphics[width=\textwidth]{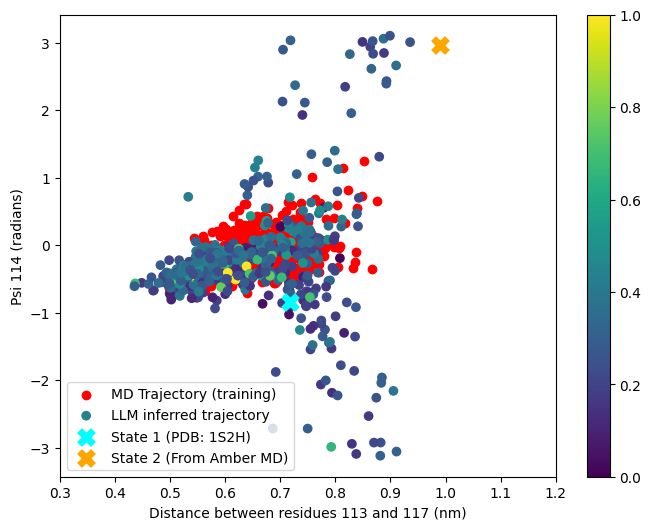}
        \caption{}
        \label{fig:mad2_scatter}
    \end{subfigure}
    \hfill
    \begin{subfigure}[b]{0.45\textwidth}
        \centering
        \includegraphics[width=\textwidth]{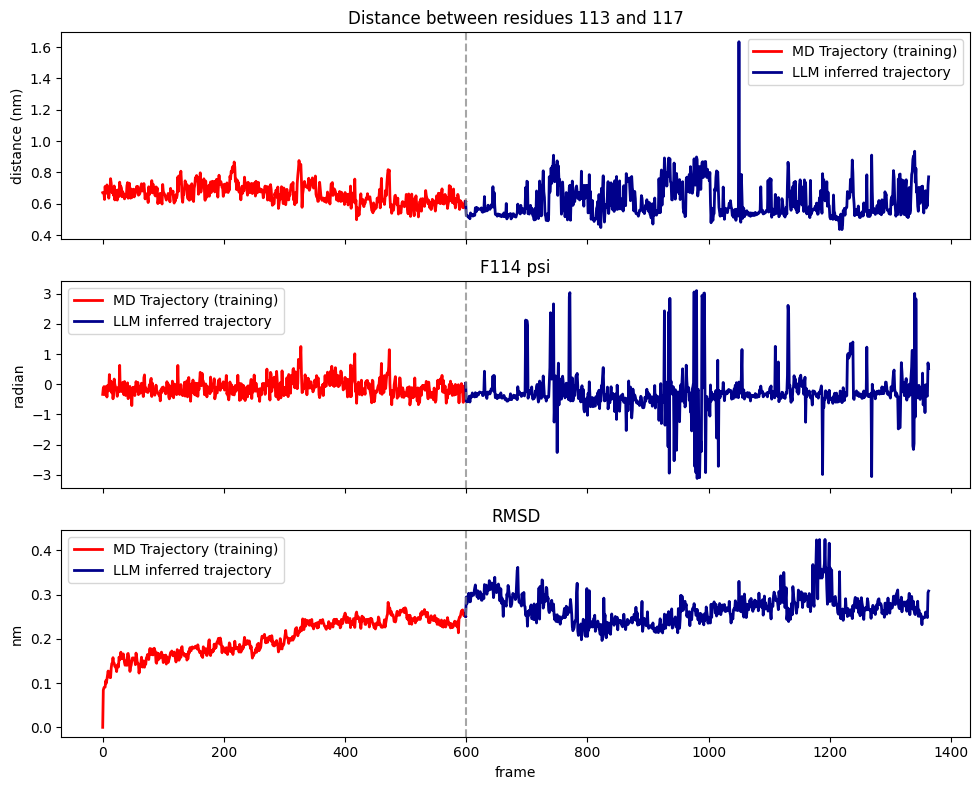}
        \caption{}
        \label{fig:mad2_ts}
    \end{subfigure}
    \caption{(a) Comparison of the regions of the conformational landscape of training data (a short MD trajectory of Mad2) and the inferred states by MD-LLM-1; the color bar corresponds to the RMSD from the reference state. (b) Time series of the characteristic distance between residues 171 and 174, the $\psi$ angle of residue 172, and RMSD starting from the training data; these parameters indicate how the MD-LLM-1 inferred trajectory extrapolates the training trajectory.}
    \label{fig:mad2}
\end{figure}
 To evaluate the generalisability of MD-LLM-1 to larger and more complex protein systems, we applied it to Mad2, a 205-residue metamorphic protein that adopts two distinct native conformations: open Mad2 (O-Mad2) and closed Mad2 (C-Mad2)\cite{JainSekhar2025}. The β7/β8 hairpin plays a key role in the conformational switching of Mad2 and undergoes substantial rearrangement during the conformational transitions, with this region experiencing transient unfolding during the O-Mad2 to C-Mad2 conversion \cite{MAPELLI2007730}. The distance between residues 171 and 174 captures the overall β-hairpin structural changes, while the ψ angle of residue 172 reports on local backbone rearrangements within this critical region. Together, these parameters provide structural reporters for tracking conformational changes between Mad2 states.
 \\
 \\
 We trained MD-LLM-1 exclusively on conformations from the C-Mad2 (closed) cluster of a Mad2 MD trajectory. Using the distance between residues 171 and 174 and the ψ angle of residue 172 as conformational coordinates, we analyzed the ability of MD-LLM-1 to discover alternative conformational states not present in the training data.
 \\
 \\
 MD-LLM-1  discovered conformations with structural parameters consistent with O-Mad2 characteristics despite being trained exclusively on the compact C-Mad2 conformations (Figure \ref{fig:mad2_scatter}). The generated conformations populated distinct regions of the conformational landscape, extending to longer distances between residues 171 and 174 characteristic of the open state. Additionally, MD-LLM-1 sampled regions between the major states, potentially representing metastable intermediates involved in the conformational conversion process.
 \\
 \\
 The discovered states showed structural features consistent with known Mad2 conformational behavior \cite{JainSekhar2025}. The model generated conformations spanning the range from the compact closed state (shorter 171-174 distances) to conformations with parameters expected for the extended open state, with residue 172 ψ angle covering the range from +0.3 radians to spans of +3 radians. This extensive ψ angle exploration is noteworthy, as β-sheet structures typically exhibit positive ψ angles (+150°) \cite{ramachandran1963stereochemistry, RICHARDSON1981167}, while the discovery of conformations with highly negative ψ angles indicates exploration of alternative backbone conformations that may correspond to the structural rearrangements associated with the transitions between the O-Mad2 and C-Mad2 states.

\section{Discussion}
\subsection{Summary of the results}
The results that we reported using the MD-LLM-1 approach provide evidence that LLMs can learn protein dynamics and infer low population states by generative modeling. Our results show that MD-LLM-1 trained on short MD trajectories can discover excited states in T4 lysozyme and alternative conformational states in Mad2, despite training exclusively on single conformational states. 
\\
\\
The MD-LLM-1 approach incorporates several features that facilitate conformational exploration. First, it generates structurally diverse yet physically plausible conformations that expand the sampled conformational space beyond the training data. Second, it learns implicit structural patterns from limited training data, effectively encoding complex relationships between residue positions without explicit physical constraints. Third, it can bypass kinetic barriers that slow down conventional sampling approaches, accessing low population states that would require long simulation times to observe through traditional methods. Fourth, the discovery capability extends to cross-state sampling, where models trained on one conformational state can predict alternative states, as shown by the bidirectional learning in T4 lysozyme and novel state discovery in Mad2.
\\
\\
The capability of the fine-tuned Mistral 7B model to produce valid protein conformations despite having no built-in structural knowledge or molecular mechanics principles is particularly noteworthy. This result suggests that LLMs can capture the underlying patterns of protein dynamics in a manner analogous to how they learn the grammar and syntax of natural languages. MD-LLM-1 effectively reads the sequential frames of protein motion and learns to infer the continuation of this motion, despite being trained on a remarkably small dataset representing only a small fraction of a conventional MD trajectory.
\\
\\
This capability suggests that with training on larger and more diverse datasets, LLMs could develop an even more comprehensive understanding of protein conformational relationships, potentially enabling direct sampling of complex free energy landscapes across broader classes of protein systems. The success of learning from limited single-state data indicates that future models trained on extensive multi-state datasets could become powerful tools for exploring protein conformational space without requiring traditional enhanced sampling techniques.

\subsection{Future directions}
Despite achieving cross-state conformation generation, including the discovery of  excited states of T4 lysozyme and alternative conformations of Mad2, the current MD-LLM-1 implementation has several limitations that motivate specific directions for future work. First, although the model reproduces structural characteristics of experimentally observed low-population states, it presently lacks explicit thermodynamic information, so relative state populations and transition probabilities cannot be derived directly from its output without performing additional energy calculations. Second, the model is not trained bidirectionally to enforce time reversibility and is not currently optimized to learn transition rates or detailed kinetic pathways. Third, the implementation remains system-specific, requiring a separate fine-tuning procedure for each protein studied. Fourth, reliance on the FoldToken structural tokenization may impose constraints on the breadth of conformational space the model can explore. 
\\
\\
Given these considerations, future work should focus on developing more generalizable models trained on diverse protein datasets,  and exploring methods to extract thermodynamic information from the learned representations. Additionally, incorporating explicit structural knowledge through specialized tokenisation schemes or attention mechanisms could further enhance the ability of MD-LLM-1 to capture physically meaningful conformational changes. We anticipate that the availability of large databases of MD trajectories\cite{liu2024dynamic, lewis2025scalable} will make it possible to overcome these limitations. 
This capability for generalization may represent the most exciting future direction for this work. Rather than training a separate model for each protein system, a single MD-LLM-1 could be trained on trajectory data from many different proteins, learning general principles of protein dynamics that apply across diverse structural contexts. This approach would mirror developments in protein structure prediction and design, where ML models have achieved remarkable generalization by learning from diverse structural data\cite{Jumper2021, abramson2024n, Krishna2024}.
\\
\\
In perspective, the application of LLMs to protein dynamics opens new opportunities for understanding and predicting biologically relevant conformational transitions. Many proteins function through complex conformational changes that often occur on timescales beyond the reach of conventional sampling methods. Generalizable approaches such as MD-LLM-1 could potentially predict these transitions with reduced computational resources, enhancing our understanding of protein function and accelerating drug discovery efforts.

\newpage
\bibliography{references}

\begin{thebibliography}{10}
\expandafter\ifx\csname url\endcsname\relax
  \def\url#1{\texttt{#1}}\fi
\expandafter\ifx\csname urlprefix\endcsname\relax\def\urlprefix{URL }\fi
\providecommand{\bibinfo}[2]{#2}
\providecommand{\eprint}[2][]{\url{#2}}

\bibitem{Benkovic2003}
\bibinfo{author}{Benkovic, S.~J.} \& \bibinfo{author}{Hammes-Schiffer, S.}
\newblock \bibinfo{title}{A perspective on enzyme catalysis}.
\newblock \emph{\bibinfo{journal}{Science}} \textbf{\bibinfo{volume}{301}}, \bibinfo{pages}{1196--1202} (\bibinfo{year}{2003}).

\bibitem{Mittermaier2006}
\bibinfo{author}{Mittermaier, A.} \& \bibinfo{author}{Kay, L.~E.}
\newblock \bibinfo{title}{New tools provide new insights in {NMR} studies of protein dynamics}.
\newblock \emph{\bibinfo{journal}{Science}} \textbf{\bibinfo{volume}{312}}, \bibinfo{pages}{224--228} (\bibinfo{year}{2006}).

\bibitem{Vendruscolo2006}
\bibinfo{author}{Vendruscolo, M.} \& \bibinfo{author}{Dobson, C.~M.}
\newblock \bibinfo{title}{Dynamic visions of enzymatic reactions}.
\newblock \emph{\bibinfo{journal}{Science}} \textbf{\bibinfo{volume}{313}}, \bibinfo{pages}{1586--1587} (\bibinfo{year}{2006}).

\bibitem{HenzlerWildman2007}
\bibinfo{author}{Henzler-Wildman, K.} \& \bibinfo{author}{Kern, D.}
\newblock \bibinfo{title}{Dynamic personalities of proteins}.
\newblock \emph{\bibinfo{journal}{Nature}} \textbf{\bibinfo{volume}{450}}, \bibinfo{pages}{964--972} (\bibinfo{year}{2007}).

\bibitem{Boehr2009}
\bibinfo{author}{Boehr, D.~D.}, \bibinfo{author}{Nussinov, R.} \& \bibinfo{author}{Wright, P.~E.}
\newblock \bibinfo{title}{The role of dynamic conformational ensembles in biomolecular recognition}.
\newblock \emph{\bibinfo{journal}{Nature Chemical Biology}} \textbf{\bibinfo{volume}{5}}, \bibinfo{pages}{789--796} (\bibinfo{year}{2009}).

\bibitem{Bonomi2016}
\bibinfo{author}{Bonomi, M.}, \bibinfo{author}{Camilloni, C.}, \bibinfo{author}{Cavalli, A.} \& \bibinfo{author}{Vendruscolo, M.}
\newblock \bibinfo{title}{Metainference: A bayesian inference method for heterogeneous systems}.
\newblock \emph{\bibinfo{journal}{Science Advances}} \textbf{\bibinfo{volume}{2}}, \bibinfo{pages}{e1501177} (\bibinfo{year}{2016}).

\bibitem{Bussi2020}
\bibinfo{author}{Bussi, G.} \& \bibinfo{author}{Laio, A.}
\newblock \bibinfo{title}{Using metadynamics to explore complex free-energy landscapes}.
\newblock \emph{\bibinfo{journal}{Nature Reviews Physics}} \textbf{\bibinfo{volume}{2}}, \bibinfo{pages}{200--212} (\bibinfo{year}{2020}).

\bibitem{Jumper2021}
\bibinfo{author}{Jumper, J.}, \bibinfo{author}{Evans, R.} \emph{et~al.}
\newblock \bibinfo{title}{Highly accurate protein structure prediction with {AlphaFold}}.
\newblock \emph{\bibinfo{journal}{Nature}} \textbf{\bibinfo{volume}{596}}, \bibinfo{pages}{583--589} (\bibinfo{year}{2021}).

\bibitem{abramson2024n}
\bibinfo{author}{Abramson, J.} \emph{et~al.}
\newblock \bibinfo{title}{Accurate structure prediction of biomolecular interactions with {AlphaFold} 3}.
\newblock \emph{\bibinfo{journal}{Nature}} \textbf{\bibinfo{volume}{630}}, \bibinfo{pages}{493--500} (\bibinfo{year}{2024}).

\bibitem{Krishna2024}
\bibinfo{author}{Krishna, R.} \emph{et~al.}
\newblock \bibinfo{title}{Generalized biomolecular modeling and design with {RoseTTAFold} all-atom}.
\newblock \emph{\bibinfo{journal}{Science}} \textbf{\bibinfo{volume}{384}}, \bibinfo{pages}{eadl2528} (\bibinfo{year}{2024}).

\bibitem{noé2019}
\bibinfo{author}{No{\'e}, F.}, \bibinfo{author}{Olsson, S.}, \bibinfo{author}{K{\"o}hler, J.} \& \bibinfo{author}{Wu, H.}
\newblock \bibinfo{title}{Boltzmann generators: Sampling equilibrium states of many-body systems with deep learning}.
\newblock \emph{\bibinfo{journal}{Science}} \textbf{\bibinfo{volume}{365}}, \bibinfo{pages}{eaaw1147} (\bibinfo{year}{2019}).

\bibitem{von2025machine}
\bibinfo{author}{von B{\"u}low, S.}, \bibinfo{author}{Tesei, G.} \& \bibinfo{author}{Lindorff-Larsen, K.}
\newblock \bibinfo{title}{Machine learning methods to study sequence--ensemble--function relationships in disordered proteins}.
\newblock \emph{\bibinfo{journal}{Current Opinion in Structural Biology}} \textbf{\bibinfo{volume}{92}}, \bibinfo{pages}{103028} (\bibinfo{year}{2025}).

\bibitem{lewis2025scalable}
\bibinfo{author}{Lewis, S.} \emph{et~al.}
\newblock \bibinfo{title}{Scalable emulation of protein equilibrium ensembles with generative deep learning}.
\newblock \emph{\bibinfo{journal}{Science}} \bibinfo{pages}{eadv9817} (\bibinfo{year}{2025}).

\bibitem{jing2024alphafold}
\bibinfo{author}{Jing, B.}, \bibinfo{author}{Berger, B.} \& \bibinfo{author}{Jaakkola, T.}
\newblock \bibinfo{title}{Alphafold meets flow matching for generating protein ensembles}.
\newblock \emph{\bibinfo{journal}{arXiv: 2402.04845}}  (\bibinfo{year}{2024}).

\bibitem{brotzakis2025alphafold}
\bibinfo{author}{Brotzakis, Z.~F.}, \bibinfo{author}{Zhang, S.}, \bibinfo{author}{Murtada, M.~H.} \& \bibinfo{author}{Vendruscolo, M.}
\newblock \bibinfo{title}{Alphafold prediction of structural ensembles of disordered proteins}.
\newblock \emph{\bibinfo{journal}{Nature Communications}} \textbf{\bibinfo{volume}{16}}, \bibinfo{pages}{1632} (\bibinfo{year}{2025}).

\bibitem{schnapka2025atomic}
\bibinfo{author}{Schnapka, V.}, \bibinfo{author}{Morozova, T.}, \bibinfo{author}{Sen, S.} \& \bibinfo{author}{Bonomi, M.}
\newblock \bibinfo{title}{Atomic resolution ensembles of intrinsically disordered and multi-domain proteins with alphafold}.
\newblock \emph{\bibinfo{journal}{bioRxiv: doi.org/10.1101/2025.06.18.660298}}  (\bibinfo{year}{2025}).

\bibitem{aranganathan2025modeling}
\bibinfo{author}{Aranganathan, A.}, \bibinfo{author}{Gu, X.}, \bibinfo{author}{Wang, D.}, \bibinfo{author}{Vani, B.~P.} \& \bibinfo{author}{Tiwary, P.}
\newblock \bibinfo{title}{Modeling {Boltzmann}-weighted structural ensembles of proteins using artificial intelligence--based methods}.
\newblock \emph{\bibinfo{journal}{Current opinion in structural biology}} \textbf{\bibinfo{volume}{91}}, \bibinfo{pages}{103000} (\bibinfo{year}{2025}).

\bibitem{alder1957phase}
\bibinfo{author}{Alder, B.~J.}, \bibinfo{author}{Wainwright, T.~E.} \emph{et~al.}
\newblock \bibinfo{title}{Phase transition for a hard sphere system}.
\newblock \emph{\bibinfo{journal}{The Journal of chemical physics}} \textbf{\bibinfo{volume}{27}}, \bibinfo{pages}{1208} (\bibinfo{year}{1957}).

\bibitem{McCammon1977}
\bibinfo{author}{McCammon, J.~A.}, \bibinfo{author}{Gelin, B.~R.} \& \bibinfo{author}{Karplus, M.}
\newblock \bibinfo{title}{Dynamics of folded proteins}.
\newblock \emph{\bibinfo{journal}{Nature}} \textbf{\bibinfo{volume}{267}}, \bibinfo{pages}{585--590} (\bibinfo{year}{1977}).

\bibitem{Karplus2002}
\bibinfo{author}{Karplus, M.} \& \bibinfo{author}{McCammon, J.~A.}
\newblock \bibinfo{title}{Molecular dynamics simulations of biomolecules}.
\newblock \emph{\bibinfo{journal}{Nature Structural \& Molecular Biology}} \textbf{\bibinfo{volume}{9}}, \bibinfo{pages}{646--652} (\bibinfo{year}{2002}).

\bibitem{car1985unified}
\bibinfo{author}{Car, R.} \& \bibinfo{author}{Parrinello, M.}
\newblock \bibinfo{title}{Unified approach for molecular dynamics and density-functional theory}.
\newblock \emph{\bibinfo{journal}{Physical review letters}} \textbf{\bibinfo{volume}{55}}, \bibinfo{pages}{2471} (\bibinfo{year}{1985}).

\bibitem{piana2011bj}
\bibinfo{author}{Piana, S.}, \bibinfo{author}{Lindorff-Larsen, K.} \& \bibinfo{author}{Shaw, D.~E.}
\newblock \bibinfo{title}{How robust are protein folding simulations with respect to force field parameterization?}
\newblock \emph{\bibinfo{journal}{Biophysical journal}} \textbf{\bibinfo{volume}{100}}, \bibinfo{pages}{L47--L49} (\bibinfo{year}{2011}).

\bibitem{vendruscolo2011protein}
\bibinfo{author}{Vendruscolo, M.} \& \bibinfo{author}{Dobson, C.~M.}
\newblock \bibinfo{title}{Protein dynamics: Moore's law in molecular biology}.
\newblock \emph{\bibinfo{journal}{Current biology}} \textbf{\bibinfo{volume}{21}}, \bibinfo{pages}{R68--R70} (\bibinfo{year}{2011}).

\bibitem{Lindorff-Larsen2005}
\bibinfo{author}{Lindorff-Larsen, K.}, \bibinfo{author}{Best, R.~B.}, \bibinfo{author}{DePristo, M.~A.} \emph{et~al.}
\newblock \bibinfo{title}{Simultaneous determination of protein structure and dynamics}.
\newblock \emph{\bibinfo{journal}{Nature}} \textbf{\bibinfo{volume}{433}}, \bibinfo{pages}{128--132} (\bibinfo{year}{2005}).

\bibitem{sarich2013e}
\bibinfo{author}{Sarich, M.}, \bibinfo{author}{Banisch, R.}, \bibinfo{author}{Hartmann, C.} \& \bibinfo{author}{Sch{\"u}tte, C.}
\newblock \bibinfo{title}{Markov state models for rare events in molecular dynamics}.
\newblock \emph{\bibinfo{journal}{Entropy}} \textbf{\bibinfo{volume}{16}}, \bibinfo{pages}{258--286} (\bibinfo{year}{2013}).

\bibitem{ray2020weighted}
\bibinfo{author}{Ray, D.} \& \bibinfo{author}{Andricioaei, I.}
\newblock \bibinfo{title}{Weighted ensemble milestoning ({WEM}): A combined approach for rare event simulations}.
\newblock \emph{\bibinfo{journal}{Journal of Chemical Physics}} \textbf{\bibinfo{volume}{152}} (\bibinfo{year}{2020}).

\bibitem{bolhuis2002transition}
\bibinfo{author}{Bolhuis, P.~G.}, \bibinfo{author}{Chandler, D.}, \bibinfo{author}{Dellago, C.} \& \bibinfo{author}{Geissler, P.~L.}
\newblock \bibinfo{title}{Transition path sampling: Throwing ropes over rough mountain passes, in the dark}.
\newblock \emph{\bibinfo{journal}{Annual review of physical chemistry}} \textbf{\bibinfo{volume}{53}}, \bibinfo{pages}{291--318} (\bibinfo{year}{2002}).

\bibitem{brotzakis2021method}
\bibinfo{author}{Brotzakis, Z.~F.}, \bibinfo{author}{Vendruscolo, M.} \& \bibinfo{author}{Bolhuis, P.~G.}
\newblock \bibinfo{title}{A method of incorporating rate constants as kinetic constraints in molecular dynamics simulations}.
\newblock \emph{\bibinfo{journal}{Proceedings of the National Academy of Sciences}} \textbf{\bibinfo{volume}{118}}, \bibinfo{pages}{e2012423118} (\bibinfo{year}{2021}).

\bibitem{Behler2007}
\bibinfo{author}{Behler, J.} \& \bibinfo{author}{Parrinello, M.}
\newblock \bibinfo{title}{Generalized neural-network representation of high-dimensional potential-energy surfaces}.
\newblock \emph{\bibinfo{journal}{Phys. Rev. Lett.}} \textbf{\bibinfo{volume}{98}}, \bibinfo{pages}{146401} (\bibinfo{year}{2007}).

\bibitem{Behler2016}
\bibinfo{author}{Behler, M.}
\newblock \bibinfo{title}{Perspective: Machine learning potentials for atomistic simulations}.
\newblock \emph{\bibinfo{journal}{Journal of Chemical Physics}} \textbf{\bibinfo{volume}{145}}, \bibinfo{pages}{170901} (\bibinfo{year}{2016}).

\bibitem{Noe2020}
\bibinfo{author}{Noé, F.}, \bibinfo{author}{Tkatchenko, A.}, \bibinfo{author}{Müller, K.-R.} \& \bibinfo{author}{Clementi, C.}
\newblock \bibinfo{title}{Machine learning for molecular simulation}.
\newblock \emph{\bibinfo{journal}{Annual Review of Physical Chemistry}} \textbf{\bibinfo{volume}{71}}, \bibinfo{pages}{361--390} (\bibinfo{year}{2020}).

\bibitem{batatia2025design}
\bibinfo{author}{Batatia, I.} \emph{et~al.}
\newblock \bibinfo{title}{The design space of e (3)-equivariant atom-centred interatomic potentials}.
\newblock \emph{\bibinfo{journal}{Nature Machine Intelligence}} \textbf{\bibinfo{volume}{7}}, \bibinfo{pages}{56--67} (\bibinfo{year}{2025}).

\bibitem{schreiner2023implicit}
\bibinfo{author}{Schreiner, M.}, \bibinfo{author}{Winther, O.} \& \bibinfo{author}{Olsson, S.}
\newblock \bibinfo{title}{Implicit transfer operator learning: Multiple time-resolution models for molecular dynamics}.
\newblock \emph{\bibinfo{journal}{arXiv: 2305.18046}}  (\bibinfo{year}{2023}).

\bibitem{bigi2025flashmd}
\bibinfo{author}{Bigi, F.}, \bibinfo{author}{Chong, S.}, \bibinfo{author}{Kristiadi, A.} \& \bibinfo{author}{Ceriotti, M.}
\newblock \bibinfo{title}{{FlashMD}: long-stride, universal prediction of molecular dynamics}.
\newblock \emph{\bibinfo{journal}{arXiv: 2505.19350}}  (\bibinfo{year}{2025}).

\bibitem{Vaswani2017}
\bibinfo{author}{Vaswani, A.} \emph{et~al.}
\newblock \bibinfo{title}{Attention is all you need}.
\newblock \emph{\bibinfo{journal}{31st Conference on Neural Information Processing Systems (NIPS)}} \textbf{\bibinfo{volume}{30}} (\bibinfo{year}{2017}).

\bibitem{zhao2023survey}
\bibinfo{author}{Zhao, W.~X.} \emph{et~al.}
\newblock \bibinfo{title}{A survey of large language models}.
\newblock \emph{\bibinfo{journal}{arXiv: 2303.18223}}  (\bibinfo{year}{2023}).

\bibitem{xiao2025protein}
\bibinfo{author}{Xiao, Y.} \emph{et~al.}
\newblock \bibinfo{title}{Protein large language models: A comprehensive survey}.
\newblock \emph{\bibinfo{journal}{arXiv: 2502.17504}}  (\bibinfo{year}{2025}).

\bibitem{lin2023evolutionary}
\bibinfo{author}{Lin, Z.} \emph{et~al.}
\newblock \bibinfo{title}{Evolutionary-scale prediction of atomic-level protein structure with a language model}.
\newblock \emph{\bibinfo{journal}{Science}} \textbf{\bibinfo{volume}{379}}, \bibinfo{pages}{1123--1130} (\bibinfo{year}{2023}).

\bibitem{madani2023large}
\bibinfo{author}{Madani, A.} \emph{et~al.}
\newblock \bibinfo{title}{Large language models generate functional protein sequences across diverse families}.
\newblock \emph{\bibinfo{journal}{Nature biotechnology}} \textbf{\bibinfo{volume}{41}}, \bibinfo{pages}{1099--1106} (\bibinfo{year}{2023}).

\bibitem{tsai2020learning}
\bibinfo{author}{Tsai, S.-T.}, \bibinfo{author}{Kuo, E.-J.} \& \bibinfo{author}{Tiwary, P.}
\newblock \bibinfo{title}{Learning molecular dynamics with simple language model built upon long short-term memory neural network}.
\newblock \emph{\bibinfo{journal}{Nature communications}} \textbf{\bibinfo{volume}{11}}, \bibinfo{pages}{5115} (\bibinfo{year}{2020}).

\bibitem{murtada2024language}
\bibinfo{author}{Murtada, M.~H.}, \bibinfo{author}{Brotzakis, Z.~F.} \& \bibinfo{author}{Vendruscolo, M.}
\newblock \bibinfo{title}{Language models for molecular dynamics}.
\newblock \emph{\bibinfo{journal}{bioRxiv: doi.org/10.1101/2024.11.25.625337}}  (\bibinfo{year}{2024}).

\bibitem{jiang2024mistral}
\bibinfo{author}{Jiang, A. Q.~{\em et al.}.}
\newblock \bibinfo{title}{Mistral {7B}}.
\newblock \emph{\bibinfo{journal}{arXiv:2310.06825}}  (\bibinfo{year}{2023}).

\bibitem{hu2021lora}
\bibinfo{author}{Hu, E.~J.} \emph{et~al.}
\newblock \bibinfo{title}{{LoRA}: Low-rank adaptation of large language models.}
\newblock \emph{\bibinfo{journal}{arXiv: 2106.09685}}  (\bibinfo{year}{2021}).

\bibitem{bouvignies2011solution}
\bibinfo{author}{Bouvignies, G.} \emph{et~al.}
\newblock \bibinfo{title}{Solution structure of a minor and transiently formed state of a {T4} lysozyme mutant}.
\newblock \emph{\bibinfo{journal}{Nature}} \textbf{\bibinfo{volume}{477}}, \bibinfo{pages}{111--114} (\bibinfo{year}{2011}).

\bibitem{JainSekhar2025}
\bibinfo{author}{Jain, S.} \& \bibinfo{author}{Sekhar, A.}
\newblock \bibinfo{title}{Transient excited states of the metamorphic protein {Mad2} and their implications for function}.
\newblock \emph{\bibinfo{journal}{Proteins}} \textbf{\bibinfo{volume}{93}}, \bibinfo{pages}{302--319} (\bibinfo{year}{2025}).

\bibitem{gao2024foldtoken4}
\bibinfo{author}{Gao, Z.}, \bibinfo{author}{Tan, C.} \& \bibinfo{author}{Li, S.~Z.}
\newblock \bibinfo{title}{Foldtoken4: Consistent \& hierarchical fold language}.
\newblock \emph{\bibinfo{journal}{bioRxiv: doi.org/10.1101/2024.08.04.606514}}  (\bibinfo{year}{2024}).

\bibitem{gao2024uniif}
\bibinfo{author}{Gao, Z.} \emph{et~al.}
\newblock \bibinfo{title}{Uniif: Unified molecule inverse folding}.
\newblock \emph{\bibinfo{journal}{Advances in Neural Information Processing Systems}} \textbf{\bibinfo{volume}{37}}, \bibinfo{pages}{135843--135860} (\bibinfo{year}{2024}).

\bibitem{unsloth_mistral7b_2024}
\bibinfo{author}{Unsloth}.
\newblock \bibinfo{title}{Mistral-7b v0.3 (4-bit) - {Hugging Face}} (\bibinfo{year}{2024}).
\newblock \urlprefix\url{https://huggingface.co/unsloth/mistral-7b-v0.3-bnb-4bit}.

\bibitem{unsloth}
\bibinfo{author}{Han, D.}, \bibinfo{author}{Han, M.} \& \bibinfo{author}{{ Unsloth team}}.
\newblock \bibinfo{title}{Unsloth} (\bibinfo{year}{2023}).
\newblock \urlprefix\url{http://github.com/unslothai/unsloth}.

\bibitem{su2024roformer}
\bibinfo{author}{Su, J.} \emph{et~al.}
\newblock \bibinfo{title}{Roformer: Enhanced transformer with rotary position embedding}.
\newblock \emph{\bibinfo{journal}{Neurocomputing}} \textbf{\bibinfo{volume}{568}}, \bibinfo{pages}{127063} (\bibinfo{year}{2024}).

\bibitem{dao2023flashattention}
\bibinfo{author}{Dao, T.}
\newblock \bibinfo{title}{Flashattention-2: Faster attention with better parallelism and work partitioning}.
\newblock \emph{\bibinfo{journal}{arXiv:2307.08691}}  (\bibinfo{year}{2023}).

\bibitem{hsu2024liger}
\bibinfo{author}{Hsu, P.-L.} \emph{et~al.}
\newblock \bibinfo{title}{Liger kernel: Efficient triton kernels for llm training}.
\newblock \emph{\bibinfo{journal}{arXiv:2410.10989}}  (\bibinfo{year}{2024}).

\bibitem{Moret2021}
\bibinfo{author}{Moret, M.}, \bibinfo{author}{Helmstädter, M.}, \bibinfo{author}{Grisoni, F.}, \bibinfo{author}{Schneider, G.} \& \bibinfo{author}{Merk, D.}
\newblock \bibinfo{title}{Beam search sampling for molecular design and intrinsic prioritization with machine intelligence}.
\newblock \emph{\bibinfo{journal}{ChemRxiv: 14153408.v1}}  (\bibinfo{year}{2021}).

\bibitem{Smith2020}
\bibinfo{author}{Smith, Z.}, \bibinfo{author}{Ravindra, P.}, \bibinfo{author}{Wang, Y.}, \bibinfo{author}{Cooley, R.} \& \bibinfo{author}{Tiwary, P.}
\newblock \bibinfo{title}{Discovering protein conformational flexibility through artificial-intelligence-aided molecular dynamics}.
\newblock \emph{\bibinfo{journal}{The Journal of Physical Chemistry B}} \textbf{\bibinfo{volume}{124}}, \bibinfo{pages}{8221--8229} (\bibinfo{year}{2020}).
\newblock \urlprefix\url{https://doi.org/10.1021/acs.jpcb.0c03985}.

\bibitem{MAPELLI2007730}
\bibinfo{author}{Mapelli, M.}, \bibinfo{author}{Massimiliano, L.}, \bibinfo{author}{Santaguida, S.} \& \bibinfo{author}{Musacchio, A.}
\newblock \bibinfo{title}{The mad2 conformational dimer: Structure and implications for the spindle assembly checkpoint}.
\newblock \emph{\bibinfo{journal}{Cell}} \textbf{\bibinfo{volume}{131}}, \bibinfo{pages}{730--743} (\bibinfo{year}{2007}).

\bibitem{ramachandran1963stereochemistry}
\bibinfo{author}{Ramachandran, G.~N.}, \bibinfo{author}{Ramakrishnan, C.} \& \bibinfo{author}{Sasisekharan, V.}
\newblock \bibinfo{title}{Stereochemistry of polypeptide chain configurations}.
\newblock \emph{\bibinfo{journal}{Journal of Molecular Biology}} \textbf{\bibinfo{volume}{7}}, \bibinfo{pages}{95--99} (\bibinfo{year}{1963}).

\bibitem{RICHARDSON1981167}
\bibinfo{author}{Richardson, J.~S.}
\newblock \bibinfo{title}{The anatomy and taxonomy of protein structure}.
\newblock \emph{\bibinfo{journal}{Advances in Protein Chemistry}} \textbf{\bibinfo{volume}{34}}, \bibinfo{pages}{167--339} (\bibinfo{year}{1981}).

\bibitem{liu2024dynamic}
\bibinfo{author}{Liu, C.} \emph{et~al.}
\newblock \bibinfo{title}{Dynamic pdb: A new dataset and a se (3) model extension by integrating dynamic behaviors and physical properties in protein structures}.
\newblock \emph{\bibinfo{journal}{arXiv: 2408.12413}}  (\bibinfo{year}{2024}).

\end{thebibliography}
\bibliographystyle{naturemag}

\end{document}